\documentclass[preprint,showpacs,preprintnumbers,amsmath,amssymb]{revtex4-1}
\usepackage{graphicx}
\usepackage{dcolumn}
\usepackage{bm}
\usepackage{epsfig}
\begin{document}
\title{Radar Cross Section of Moving Objects}

\author{H. Gholizade}\altaffiliation {
gholizade@ut.ac.ir} \affiliation{Engineering Science Department,
University of Tehran, Tehran, Iran.}
\begin{abstract}
I investigate the effects of movement on radar cross section
calculations. The results show that relativistic effects (the constant
velocity case) can change the RCS of moving targets by changing the
incident plane wave field vectors. As in the Doppler effect, the
changes in the fields are proportional to $\frac{v}{c}$. For accelerated
objects, using the Newtonian equations of motion yields an effective
electric field (or effective current density) on the object due to the finite mass of the conducting electrons. The results
indicate that the magnetic moment of an accelerated object is different from that of an
un-accelerated object, and this difference can change the RCS of the object. Results for moving sphere and non-uniformly rotating sphere are given and compared with static (\textbf{v}=0) case.
\end{abstract}

\pacs{} \maketitle
\section{\label{sec:level1}Introduction}
After Einstein's theory of relativity, we know that some quantities,
such as four-vectors, have a norm which is invariant under Lorentz transformations.
Maxwell's equations have a relativistic nature and we can calculate
the relativistic effects on their solutions. The most
well known effect is the Doppler effect. The Doppler effect can be described by
the phase invariance of the wave equation. The relativistic Doppler effect has
new consequences, indescribable in the Newtonian point of view. For
example, the transverse Doppler effect, i.e., when the source velocity and
the propagation direction are orthogonal. In the non-relativistic Doppler effect,
the frequency change is proportional to the cosine of the relative angle between
the directions of velocity and propagation. But in the relativistic case, we know that
if $\cos(\theta)=0$ then we have a frequency shift \cite{relativity}. This effect is different from the
rotational Doppler effect, because a rotating frame is not an inertial frame, so we can not use the special theory of relativity to describe its physics.
In the case of rotating frames, we can use "Franklin" transformation \cite{noori}. In relativity the linear velocity of rotating frame is $v(r) = c \tanh(\frac{r \omega}{c})$ against $v(r)=r \omega$. Several authors discuss about relativistic corrections of EM fields for both constant velocity and rotating frames \cite{m1,m2,r1,r2,r3,r4,r5,r6,r7,r8,r9,r10}.
At first sight, it seems that relativistic effects can not be detected, but we know that relativistic
effects in engineering can be considered \cite{halliday}. In this paper I focus on
both relativistic effects (constant velocity) and
non-relativistic effects (accelerated objects) on the scattering
process that can change the differential scattering cross section of
the scatterer. (The differential scattering cross section is
proportional to the radar cross section in engineering \cite{jakson} by
the factor $4 \pi$.) I ignore the changes in frequency due to target
movement (the Doppler effect), changes in the incident angles, charge and
current densities, target and source relative distances, and only
investigate the changes in the fields and their effects on the scattering
cross section.
In this paper I focus on multi- static RCS (differential cross section), and express it as a perturbative series ($\sigma = \sigma_0(1+\delta)$) up to $\frac{v}{c}$ order.
I also solve equations for simple case (moving sphere) and compar the results with static case ($\textbf{v}=0$).
In the relativistic case, I use standard Lorentz transformations to find the observed fields in the different frames.
This process can be found in most textbooks \cite{jakson, relativity}. In the case of accelerated motion,
I use the effect of the conducting electrons' finite mass (the Stewart--Tolman effect) \cite{electrodynamics landau} and its effects on the current distribution. The result of Stewart--Tolman effect can be collected as an effected magnetic moment \cite{electrodynamics landau}. The scattering cross section in Rayleigh region was calculated for non- uniformly rotating sphere.
First, I give a short review of the special theory of
relativity in Section II. Section III discusses the RCS of moving objects.
Section IV investigates the effects of target acceleration on
the RCS, and the results for a simple model (Rayleigh
scattering for a conducting sphere) are given.
\section{Relativity in Electrodynamics}\label{sec1}
Let $K$ and $K'$ be inertial frames and suppose that $K'$ moves with
constant velocity $\textbf{v}$ relative to $K$. The observer $O$
(i.e., in frame $K$) measures the electromagnetic fields $\textbf{H}$,
$\textbf{E}$, and observer $O'$ (in frame $K'$) also measures the same fields as
$\textbf{H}'$ and $\textbf{E}'$. To obtain the relation between
the fields in the two frames, we can use the field strength tensor \cite{relativity}:
\begin{equation}
  F^{\mu \nu}= \partial^\mu A^\nu - \partial^\nu A^\mu.
  \label{eq1}
\end{equation}
Here, $A^\mu$ $(\mu =0,1,2,3)$ is the potential four-vector.
The transformed tensor in the frame $K'$ can be written as
\begin{equation}
  F'^{\mu \nu}=\frac{\partial x'^\mu}{\partial x^\alpha}\frac{\partial x'^\nu}{\partial x^\beta} F^{\alpha
  \beta}.
  \label{eq2}
\end{equation}
Using the Lorentz transformation for $x'^\mu$ we find the
following relation between the fields in the two frames \cite{relativity, jakson}:
\begin{eqnarray}
  \textbf{E}'=\gamma (\textbf{E}+\beta\times\textbf{B})-\frac{\gamma^2}{\gamma+1}\beta(\beta.\textbf{E}),\\ \nonumber
  \textbf{B}'=\gamma
  (\textbf{B}-\beta\times\textbf{E})-\frac{\gamma^2}{\gamma+1}\beta(\beta.\textbf{B}).
  \label{eq3}
\end{eqnarray}
If we keep only terms of order $\frac{v}{c}$, then \cite{relativity, jakson}
($\beta=\frac{\textbf{v}}{c}, \gamma^{-1}=\sqrt{1-\beta^2}$):
\begin{eqnarray}
\textbf{E}'=\textbf{E}+\beta\times\textbf{B},\\ \nonumber
\textbf{B}'=\textbf{B}-\beta\times\textbf{E}. \label{eq4}
\end{eqnarray}
Therefore, as the moving observer sees a different frequency, they measure
different fields also. The order of this difference is proportional
to $\frac{v}{c}$, as in the  Doppler effect.
\section{The RCS of a Moving Conductor}
When a conductor (or dielectric) moves with respect to a source, then, according to Section \ref{sec1}, the electromagnetic fields change in amplitude, direction, and frequency.
These changes are very small compared to the original fields,
 so we can expand the scattered fields in terms of $\frac{v}{c}$. According to standard text books \cite{jakson}, the scattered field can represented by
\begin{eqnarray}
    \textbf{H}^{'s}(r)=\frac{1}{4 \pi} \int_{s'}[(\hat{n}'\times \textbf{H}^{'T})\times \nabla',
    \psi_0]ds',\\
     \textbf{E}^{'s}(r)=\frac{1}{4 \pi} \int_{s'}[\imath \omega'\mu'(\hat{n}'\times \textbf{H}^{'T})\psi_0+(\hat{n}'\cdot \textbf{E}^{'T})\times \nabla' \psi_0]ds',\label{eq5}
\end{eqnarray}
where $\psi_0$ is the free space Green's function \cite{rcs handbook}:
\begin{equation}
   \psi_0 = \frac{e^{\imath k |\textbf{r}-\textbf{r}'|}}{|\textbf{r}-\textbf{r}'|}.\nonumber
\end{equation}
$\textbf{H}^{'T}$ and $\textbf{E}^{'T}$ are the total electromagnetic fields at the surface of the scatterer,
\begin{eqnarray}\label{eq6}
\textbf{H}^{'T} = \textbf{H}^{'i}+\textbf{H}^{'s},\\
\textbf{E}^{'T} = \textbf{E}^{'i}+\textbf{E}^{'s}.
\end{eqnarray}
The $\textbf{H}^{0i}$ is the radiating field in frame $K$ and $\textbf{H}^{'i}$ ($\textbf{H}^{'s}$) is the incoming (scattered) field in frame $K'$.
The relation between $\textbf{H}^{'0}$ and $\textbf{H}^{i}$ is
\begin{eqnarray}\label{eq7}
\textbf{H}^{'i} = \textbf{H}^{0i}-\delta \textbf{H}',\\ \nonumber
\delta \textbf{H}'=\beta \times \textbf{E}^i.
\end{eqnarray}
Inserting these equations into (\ref{eq5}), we obtain
\begin{eqnarray}\label{eq8}
    \textbf{H}^{'s}(r)=\frac{1}{4 \pi} \int_{s'}[(\hat{n}'\times (\textbf{H}^{'0T}+ \delta \textbf{H}'))\times \nabla'
    \psi_0]ds'.\\ \nonumber
    \textbf{H}^{'s}(r)=\textbf{H}^{0s}(r)-\delta \textbf{H}^{'s},\\
    \delta \textbf{H}^{'s}=\frac{1}{4 \pi} \int_{s'}[(\hat{n}'\times \delta \textbf{H}')\times \nabla'
    \psi_0]ds'.
\end{eqnarray}
Here, $\textbf{H}^{0s}(r)$ is the scattering field when the relative
velocity of $K$ and $K'$ is zero. It must be mentioned here that I
ignore the relativistic effects that change the normal vector, the
free space Green's function, and $ds'$, and assume that all these
quantities are the same as when $K'$ is at rest with respect to $K$.
Using Lorenz transformation by inserting $\textbf{v}\rightarrow
\textbf{-v}$, we obtain fields in $K$ frame. Keeping only terms of
order $\frac{v}{c}$ then the scattered field in frame $K$ is:
\begin{eqnarray}\label{hsk}
\textbf{H}^{s} = \textbf{H}^{0s}+\delta \textbf{H}^s,\\ \nonumber
\delta \textbf{H}=\delta \textbf{H}^{'s}-\beta \times \textbf{E}^s.
\end{eqnarray}
The radar cross section (RCS) is \cite{rcs handbook}
\begin{equation}\label{eq9}
\sigma= 4 \pi r^2 \frac{{\textbf{H}^s}^*.\textbf{H}^s}{{\textbf{H}^i}^*.\textbf{H}^i}.
\end{equation}
By expanding the denominator,
\begin{equation}\label{eq10}
\sigma= 4 \pi r^2({\textbf{H}^s}^*.\textbf{H}^s)(\sum_{l=0}^{\infty}\sum_{m=-l}^{l}\frac{4\pi}{2l+1}
\frac{|{\delta H}^i|^l}{|H^{0i}|^{l+1}}Y'^*_{l,m}(\theta',\phi')Y_{l,m}(\theta,\phi))^2.\\
\end{equation}
The velocity of the moving object is very small compared to the velocity of light, therefore we can write
\begin{equation}
\delta\textbf{ H}^i \ll  \textbf{H}^i.\nonumber
\end{equation}
In this situation, we may keep only the first and second terms in expansion.
\begin{eqnarray}\label{eq11}
      \sigma \simeq 4 \pi r^2({\textbf{H}^s}^*.\textbf{H}^s)^2 (\frac{4\pi}{H^i}Y'^*_{0,0}(\theta',\phi')Y_{0,0}(\theta,\phi)+\frac{4\pi}{3}
      \frac{\delta H^i}{{H^{0i}}^2}\sum_{m=-1}^{1}[ Y'^*_{1,m}(\theta',\phi')Y_{1,m}(\theta,\phi)])^2.
\end{eqnarray}
\begin{eqnarray}\label{eq12}
\sigma \simeq  \sigma^0+\frac{8 \pi}{3} \sigma^0 \frac{\delta
H^i}{H^{0i}}\sum_{m=-1}^{1}
Y'^*_{1,m}(\theta',\phi')Y_{1,m}(\theta,\phi)+2\frac{\textbf{H}^{0s}.\delta\textbf{H}^{s}}{{H^{0i}}^2},
\end{eqnarray}
where in above equation we have
\begin{eqnarray}
    \textbf{H}^s = \textbf{H}^{0s}+ \delta \textbf{ H}^s,\\ \nonumber
    \sigma^0= 4 \pi r^2 \frac{{\textbf{H}^{0s}}^*.\textbf{H}^{0s}}{{\textbf{H}^{0i}}^*.\textbf{H}^{0i}}.\nonumber
\end{eqnarray}
In equations (\ref{eq11}, \ref{eq12}) $\theta',\phi'$
($\theta,\phi$) are the angles of $\delta\textbf{ H}^i$ ($\textbf{
H}^i$) in spherical polar coordinates. It is evident that
$\theta',\phi'$ contains the information about the direction of
motion. The third term in (\ref{eq12}) can written as
\begin{eqnarray}\label{eq13}
2\frac{{H^{0s}}^2}{{H^{0i}}^2}\frac{\delta H^{s}}{H^{0s}}
\cos(\gamma)\\ \nonumber =2 \sigma^0\frac{\delta
H^{s}}{H^{0s}}\cos(\gamma).
\end{eqnarray}
The order of this correction is proportional to $\frac{v}{c}$. It
must be noted that the boundary condition for the total fields
$\textbf{H}^T$ and $\textbf{H}^s$ must be satisfied \cite{tolman, m1}.
Collecting results, we can write following result for RCS up to
order $\frac{v}{c}$ as:
\begin{equation}\label{eq13-1}
\sigma=\sigma^0\left(1+\frac{8 \pi}{3} \frac{\delta
H^i}{H^{0i}}\sum_{m=-1}^{1}
Y'^*_{1,m}(\theta',\phi')Y_{1,m}(\theta,\phi)+2\frac{\delta
H^{s}}{H^{0s}}\cos(\gamma)\right).
\end{equation}
\begin{figure}[htb]
\epsfxsize=7cm \epsfbox{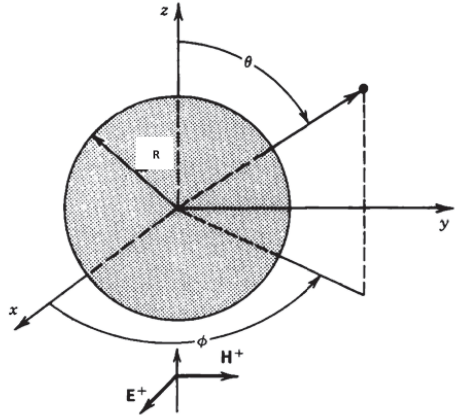} \caption{Geometry of incident plane wave and sphere.}\label{fig3}
\end{figure}
Equation (\ref{eq13-1}) has convenience form for relativistic
corrections, because all quantities are represented as static
($v=0$) fields.
 As an example I assume a sphere with radius R and moving with velocity $\textbf{v} =v \widehat{x}$. For simplicity I set (as shown in figure (\ref{fig3})):
\begin{equation}
\textbf{E}^i=E_0 \widehat{x} e^{-\imath k r \cos(\theta)}.
\end{equation}
In this case we have:
\begin{equation}
\beta \times \textbf{E}^i = 0.
\end{equation}
By above simplification; the only corrections in RCS are due to last term in equation (\ref{hsk}) i.e. $- \beta \times \textbf{E}^s$.
According to standard text books, \cite{jakson, balanis,rcs handbook}, the scattered fields at far field limit for sphere can be written as:

\begin{figure}[htb]
\epsfxsize=7cm \epsfbox{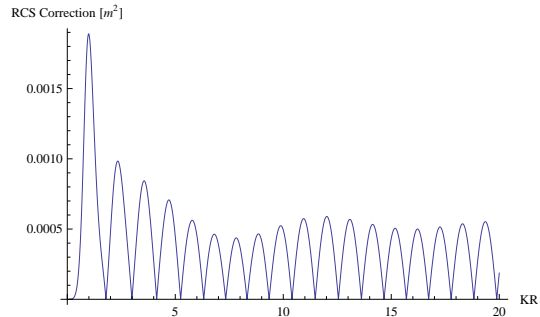} \caption{Relativistic corrections for mono static RCS as a function of $kR$ (R is sphere radius) for $\beta = 10^{-5}$. This figure can be compared with figure (\ref{fig5}) that indicate the mono-static RCS of same sphere. The ratio of relativistic corrections and static RCS is (approximately) of order $10^{-4}$.}\label{fig2}
\end{figure}

\begin{eqnarray}
E_r^s\cong 0,\\
E_{\theta}^s\cong \imath E_0 \frac{\exp(-\imath k r)}{k r}f(\theta,\phi),\\
E_{\phi}^s\cong \imath E_0 \frac{\exp(-\imath k r)}{k r}g(\theta,\phi).
\end{eqnarray}
With definitions:
\begin{eqnarray}
f(\theta,\phi)=\cos(\phi)\sum_{n=1}^{\infty}\imath^n\left\{b_n \sin(\theta)\frac{\partial P_n^1(\cos(\theta))}{\partial \cos(\theta)}-c_n \frac{P_n^1(\cos(\theta))}{\sin(\theta)}\right\},\\ \nonumber
g(\theta,\phi)=\sin(\phi)\sum_{n=1}^{\infty}\imath^n\left\{b_n \frac{P_n^1(\cos(\theta))}{\sin(\theta)}-c_n \sin(\theta)\frac{\partial P_n^1(\cos(\theta))}{\partial \cos(\theta)}\right\},\\ \nonumber
a_n=i^{-n}\frac{(2 n +1)}{n(n+1)},\\ \nonumber
b_n=-a_n\frac{((1+n) j_n(k R)-k R j_{1+n}(k R))}{((1+n) h^{(2)}_n(k R)- k R h^{(2)}_{1+n}(k R))},\\ \nonumber
c_b=-a_n \frac{j_n(k R)}{h^{(2)}_n(k R)}.
\end{eqnarray}
the results are shown in figures (\ref{fig3},\ref{fig2}) for sphere moving with $v = 10^3 m/s$. In figure (\ref{fig2}) the results are shown for relativistic corrections for mono-static RCS and we can compare it with figure (\ref{fig5}) that shows mono- static RCS for same geometry. The order of corrections is ($-30 dBm\longleftrightarrow-40 dBm$). Comparing this results with mono-static RCS of sphere ($1dBm \longleftrightarrow 10dBm$) shows that the relativistic corrections and static RCS ratio approximately of order (-40 dBm).
\begin{figure}[htb]
\epsfxsize=7cm \epsfbox{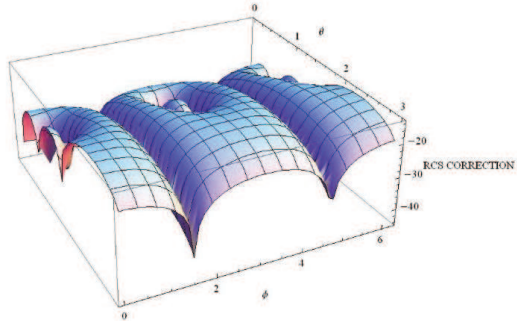} \caption{The RCS relativistic corrections as a function of $\theta, \phi$ and $kR=1$ , $\beta = 10^{-5}$.}\label{fig4}
\end{figure}
This result is predictable, because we expect that the relativistic corrections in first order are of order $\beta= \frac{v}{c}$.
\begin{figure}[htb]
\epsfxsize=7cm \epsfbox{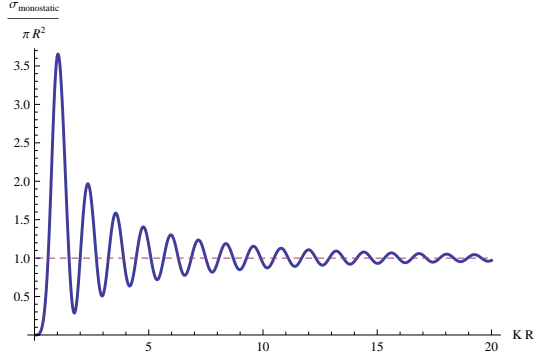} \caption{Mono- static RCS for sphere as a function of $kR$ .}\label{fig5}
\end{figure}

\section{The Effect of Target Acceleration on the RCS}
In the previous section, I only studied the case of constant velocity (the special
theory of relativity) and its effects on the scattering of the electromagnetic waves.
In this section, I will discuss the case of acceleration of the target
and the possible effects of this acceleration. All the relations are based on Newtonian mechanics and I
don't discuss accelerated frames in relativity. The effect of
acceleration can be taken into account by assuming inertial forces on
the conduction electrons \cite{electrodynamics landau}. This effective
force on an electron is $-m_e \dot{\textbf{v}}$ and this is equivalent
to an electric field $ \frac{-m_e\dot{\textbf{v}}}{e}$
\cite{electrodynamics landau}.
\begin{equation}\label{eq14}
\textbf{E}'=\textbf{E}+\frac{m_e\dot{\textbf{v}}}{e}.
\end{equation}
If we assume that $\textbf{v}=\textbf{u}+\overrightarrow{\Omega}
\times \textbf{r}$, where $\textbf{u}$ is the linear velocity and
$\overrightarrow{\Omega}$ is the angular velocity, then we have
($\mu_r=1$)
\begin{eqnarray}\label{eq15}
\nabla \times \textbf{E}'= -\mu_0 \frac{\partial
\textbf{H}'}{\partial t}.
\end{eqnarray}
In (\ref{eq15}), $\textbf{H}'$ is \cite{electrodynamics
landau}
\begin{equation}\label{eq16}
\textbf{H}'=\textbf{H}-2 m_e \frac{\overrightarrow{\Omega}}{e \mu_0}.
\end{equation}
It can be shown \cite{electrodynamics landau} that solving the
variable magnetic field near a non-uniformly rotating body is
equivalent to solving the problem of a body at rest in a uniform
external magnetic field:
\begin{equation}\label{eq17}
\textbf{H}_{equ}=2 m_e \frac{\overrightarrow{\Omega}}{e \mu_0}.
\end{equation}
According to the above equations, when a conducting sphere with $\mu_r 1$ lies in a uniform periodic external field, its magnetic moment is $V
\alpha \textbf{H}$; where $V$ is the volume of the sphere, $a$ is the radius of the sphere,
and \cite{electrodynamics landau}
\begin{eqnarray}
\alpha=-\frac{3}{2}[1-\frac{3}{a^2 \kappa^2}+\frac{3}{a \kappa}\cot(a\kappa)]\\
\nonumber
\kappa=\frac{1+\imath}{\delta}.\nonumber
\end{eqnarray}
Here, $\delta$ is the penetration depth. In the case of a non-uniformly
rotating sphere, we have ($\delta \ll a$)\cite{electrodynamics landau}:
\begin{equation}\label{eq18}
\textbf{m}=\frac{m_e a^5 \varsigma}{15 e \mu_0}\frac{d
\overrightarrow{\Omega}}{dt}.
\end{equation}
In (\ref{eq18}), $\varsigma$ is the conductivity of the sphere.
If the incident wavelength is very large compared to the radius of the sphere (Rayleigh scattering), then we can use the following relations between the  incident and the scattered fields \cite{jakson}:
\begin{align}
\textbf{E}_{inc}&=\textbf{d}_0 E_0 \exp(\imath k
\textbf{n}_0.\textbf{r}),\\ \nonumber \textbf{E}_{sc}&=\frac{k^2
\exp(\imath k r)}{4 \pi \varepsilon_0 r}[ (\textbf{n}\times
\textbf{p})\times \textbf{n}]-\frac{\mu_0\exp(\imath k r)}{4 \pi r
c}[\hat{\phi} \sin(\theta)\frac{\partial^2 m(t)}{\partial t^2}].
\end{align}
In the above equations, $\textbf{d}_0$, $\textbf{n}_0$
($\textbf{d}$, $\textbf{n}$) are the incident (scattered) electric field
polarization and plane wave propagation direction, respectively
($\textbf{p}$ is the electric dipole moment of the sphere and I assumed that
$\textbf{m}(t)= m(t) \hat{k}$.). The first term represents the
electric dipole scattering and the second term is the electric field radiating from
the time dependent magnetic moment. In this case, the differential
scattering cross section ($\frac{\sigma}{4 \pi}$) becomes
\begin{equation}\label{eq20}
\frac{d\sigma}{d \Omega}=|\frac{k^2}{4 \pi \varepsilon_0 E_0}(\textbf{p}.\textbf{d}^*)\textbf{d}+\frac{\mu_0}{4 \pi E_0
c}\hat{\phi} \sin(\theta)\frac{\partial^2 m(t)}{\partial t^2}|^2.
\end{equation}
\begin{figure}[htb]
\epsfxsize=7cm \epsfbox{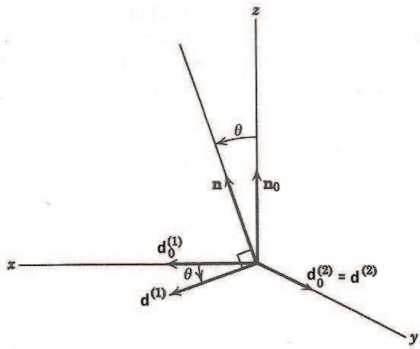} \caption{The geometry of the incident and scattered fields.}\label{fig1}
\end{figure}
For the special definition of $\textbf{d}_0$, $\textbf{n}_0$ ,$\textbf{d}$, $\textbf{n}$, as shown in Figure (\ref{fig1}), we have
\begin{eqnarray}\label{eq21}
\frac{d\sigma_{\|}}{d \Omega}=\frac{k^4 a^6}{2} \cos^2(\theta)+\frac{1}{2}(\frac{\mu_0 \ddot{m}(t)\sin(\theta)}{4 \pi c E_0})^2\\ \nonumber
\frac{d\sigma_{\perp}}{d \Omega}=\frac{k^4 a^6}{2} +\frac{1}{2}(\frac{\mu_0 \ddot{m}(t)\sin(\theta)}{4 \pi c E_0})^2.
\end{eqnarray}
It is evident that the resulting scattering cross section is completely different from the  non-rotating case. An unusual result
 appears in (\ref{eq21}), and the scattering cross section depends on the incident electric field strength. This is an indication
 of the presence of nonlinear effects in the scattering process.
\section{Results and Discussion}
In the present paper, I investigated the possible effects of target
movement on the scattering cross section, i.e., the RCS. The theoretical
calculations show that when the target has a velocity with respect to
the source, then the incident plane wave parameters change. These changes
include the frequency (the Doppler effect), the incident angles, and the field
vectors ($\textbf{E}$ , $\textbf{H}$) (in magnitude and direction). I
only studied the changes in the fields and the results showed that a moving
target RCS is different from the motionless (standing or static) case. This difference arises
from changes in the electric and magnetic fields due to relativistic
effects (constant velocity case) and non-relativistic effects
(an accelerated object). When the object moves with constant velocity
then, according to the special theory of relativity, the observer that moves
with the object measures different field vectors and frequencies. The
changes in the frequency can explained by the phase invariance of the energy-momentum four-vector ($h\nu/c, \textbf{p}$) under Lorentz transformations
for the incident photons. The field changes can be obtained from the field
strength tensor transformation. As with  the Doppler effect, the order magnitude of
the changes in the field is proportional to $\frac{v}{c}$. This means that
the moving corrections are very small compared to the static case,
as with the  Doppler effect. When the incident frequency is on the order of MHz--GHz,
we can measure the Doppler frequency, but in the case of fields, the
strength of the field is not large enough, and so the detection of relativistic
effects becomes difficult. For accelerated frames, I used the
results  of Newtonian mechanics and added an inertia force to the conducting
electrons' equation of motion. This is equivalent to changes in
the effective electric field (or current density). When we use the
effective electric field or effective current density, it can be shown
\cite{electrodynamics landau} that the magnetic moment of the accelerated
object depends on the angular acceleration (not the total acceleration).
This means that the magnetic moment of a non-uniformly rotating sphere is
different from that of a non-rotating one. Changes in the magnetic moment
directly change the RCS, and the
differential scattering cross section was calculated in the Rayleigh scattering regime. In this paper, I used the far-field limit of the radiating electromagnetic fields with a time-dependent magnetic moment, and didn't solve Maxwell's equations. For a non-uniformly rotating sphere, the solution outside of the sphere is similar to that for a non-rotating one, i.e., $\triangle \textbf{H}'=0$. On the surface of the sphere, $\textbf{H}'$ is continuous, and outside the sphere, at infinity, $\textbf{H}'$ becomes $-\textbf{H}_{equ}$.
The effect  of a non-uniform rotation in a ring appears as an e.m.f and is known as the \textit{Stewart--Tolman} effect.
The other possible effects of this movement, such as small surface deformations and thermal effects, were not included in this paper. These effects generally depend on the shape and structure of the moving object and can be investigated as independent parameters in RCS calculation.
\section{Acknowledgments}
I would like to thank the Research Council of the University of
Tehran and the Institute for Research and Planning in Higher
Education for financial support and grants provided under
contract No. 138-569.

\end{document}